\nonstopmode

\documentclass[aps,prl,twocolumn,showpacs]{revtex4}
\usepackage{graphicx}
\begin{document}
\title{Experimental Phonon Band Structure of Graphene using C$^{12}$ and C$^{13}$ Isotopes}
\author{S. Bernard}
\author{E. Whiteway}
\author{V. Yu}
\author{D.G. Austing$^*$}
\author{M. Hilke}
\affiliation{Department of Physics, McGill University, Montr\'eal, Canada H3A 2T8}
\date{\today}

\begin{abstract}

Using very uniform large scale chemical vapor deposition grown graphene transferred onto silicon, we were able to identify 15 distinct Raman lines associated with graphene monolayers. This was possible thanks to a combination of different carbon isotopes and different Raman laser energies and extensive averaging without increasing the laser power. This allowed us to obtain a detailed experimental phonon dispersion relation for many points in the Brillouin zone. We further identified a D+D' peak corresponding to a double phonon process involving both an inter- and intra-valley phonon.

\end{abstract}
\maketitle

Graphene has attracted a considerable amount of attention due to the ease in isolating a single sheet of graphite via mechanical exfoliation \cite{novo04,novo05}. Despite the fact that it is one atom thick, exfoliated graphene has shown extraordinary electronic, vibrational and optical properties that can be used as a novel material for many potential applications. Its unique electronic band structure constitutes its most noteworthy property, with the existence of two degenerate Dirac cones \cite{Wallace47}, which leads to two degenerate valleys (K and K'). Similarly, its vibrational properties, characterized by its phonon band structure is also composed of two degenerate Dirac cones for the out-of-plane modes. These out-of-plane modes, however, do not play any significant role in relation to graphene's electronic properties, in contrast to the in-plane modes, which can couple to the electronic modes. The four in-plane modes are composed of the transverse (TA) and longitudinal (LA) acoustic modes and the transverse (TO) and longitudinal (LO) optical modes, which are responsible, for example, for the very high thermal conductivities observed in graphene \cite{thermo}.

The most striking example of the interplay between optical, vibrational and electronic properties, can be found in inelastic light scattering (Raman). The photo-excited electrons, which live in the Dirac cones, can inelastically scatter with the vibrational modes and elastically with defects, either within one valley or between valleys and back. This leads to a thorough probing of the available phase space, for both electrons and phonons. The goal in this letter is to use this rich photon-phonon-electron interplay, in order to deduce the detailed phonon properties of graphene, including the in-plane phonon band structure, using the well characterized photon and electron states. While for graphite, other methods exist to experimentally characterize the phonon band structure, such as inelastic X-ray scattering \cite{IXS}, they do not give enough signal with graphene, because of the two-dimensional nature. Hence, only a very limited number of experimental data points for the phonon band structure were obtained using Raman spectroscopy on exfoliated graphene \cite{Mafra07}.

In order to gain a more complete picture with sufficient precision and resolution it is important to avoid finite size effects with different edge configurations, which can influence both the electronic and vibrational properties of graphene. Moreover, the substrate also plays an important role in some of the properties. Hence in order to both eliminate substrate effects and to probe large areas, we undertook to study Raman scattering for large scale chemical vapor deposition (CVD) grown graphene using two different isotopes (C$^{12}$ and C$^{13}$) so that we can effectively exclude and subtract the substrate contributions, since a heavier mass downshifts only the vibrational properties, while keeping all other properties the same. The combination of C$^{12}$ and C$^{13}$ was also used in CVD-grown graphene to investigate the growth mechanism on Ni and Cu \cite{li092} and to study doping effects when prepared as a bilayer graphene \cite{kalbac11}.

For this work we grew graphene monolayers by CVD of hydrocarbons on 25 $\mu$m-thick commercial Cu foils. The Cu foil is first acid-treated for 10 mins using acetic acid and then washed thoroughly with de-ionized water. Graphene growth is realized in conditions similar to Li and others \cite{li09,bae10,Yu11}, but using a vertical quartz tube. We used two isotopic methane sources to grow our graphene: a 99.99\% pure C$^{12}$-methane and a 99.9\% C$^{13}$-methane (CLM-3590-1) from Cambridge Isotopes Laboratories, Inc. Each gas was used to grow graphene on a copper substrate. The graphene is grown at 1025$^\circ$C in 0.5 Torr, with a 4 sccm H$_2$ flow and a 40 sccm CH$_4$ flow for 30 minutes. The methane flow is stopped while the hydrogen flow is kept on during the cooling process. The graphene film was transferred onto n$^{++}$-doped Si wafer with an oxide thickness of 285nm. This is realized by spin coating the graphene-Cu with PMMA, followed by an etching of the Cu foil in an oxidizing solution of 0.1M ammonium persulfate((NH$_4$)$_2$S$_2$O$_8$).

Here we performed Raman spectroscopy experiments using different laser energies. While there are both Stokes and anti-Stokes Raman shifts, we only focused on the Stokes lines in this letter, since they are more pronounced. Hence, we will only consider processes where phonons are emitted by electron-phonon scattering.

\begin{figure}[ptb]
\begin{center}
\includegraphics[width=3.0in]{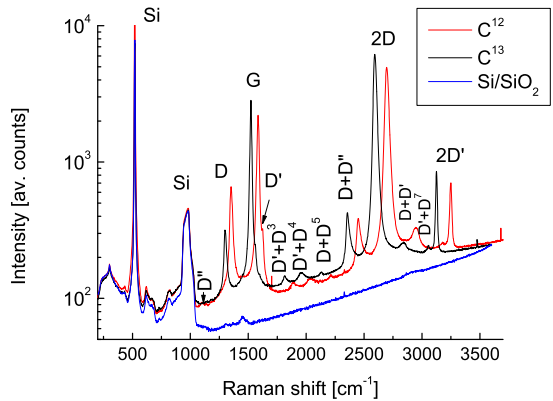}
\caption{Photon counts averaged over about 60 Raman spectra at different locations of the sample as a function of the Raman shift for C$^{13}$-graphene and C$^{12}$-graphene on SiO$_2$ covered Si substrates. The peaks labeled Si are attributed to the Si/SiO$_2$ substrate.}%
\label{spectrum}%
\end{center}
\end{figure}

The Raman spectrum for the C$^{13}$-graphene sample is expected to shift downward from the C$^{12}$-graphene by a factor of $\sqrt{13/12}$ (the square root of the ratio of the atomic masses). The Raman spectra were obtained by taking the spectra over many different locations (between 50-70) of the same sample (of cm$^2$ size). This allows us to confirm its uniformity and to perform a large configurational average to better resolve very weak peaks, such as the D'+D$^3$, D'+D$^4$, and D+D$^5$ Raman lines, which were observed recently \cite{D3D4D5}. In addition, we identified more new weak Raman lines, including D", D$^3$, D$^4$, D$^5$, D$^6$, D+D' and D'+D$^7$, some of which were expected theoretically \cite{Mauri11}. The attribution of these new peaks along with their peak positions at different laser energies enables us to map out a detailed experimental phonon band structure.

\begin{figure}[ptb]
\begin{center}
\hbox{\includegraphics[width=2in]{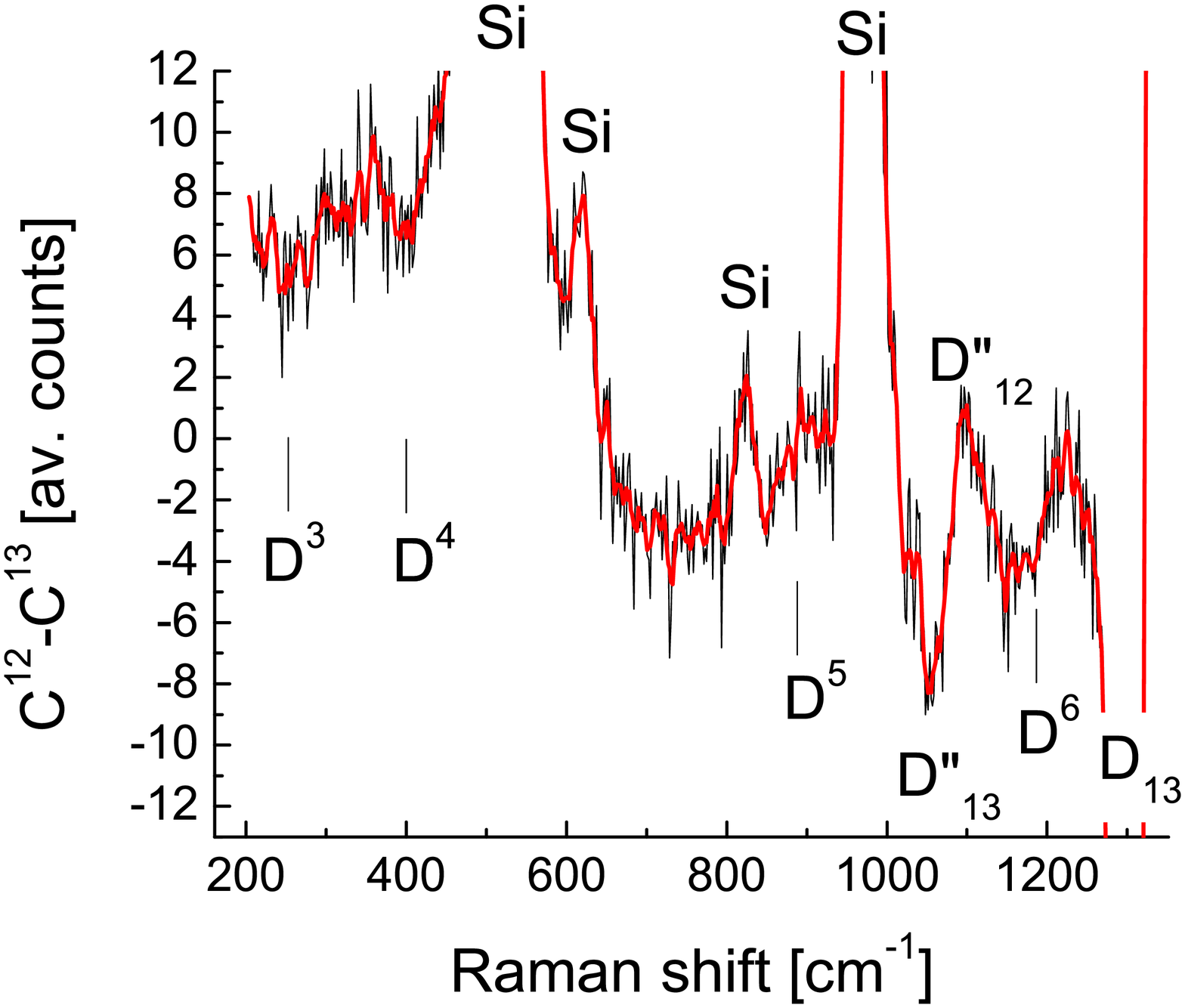}\hspace*{-1cm}\includegraphics[width=2in]{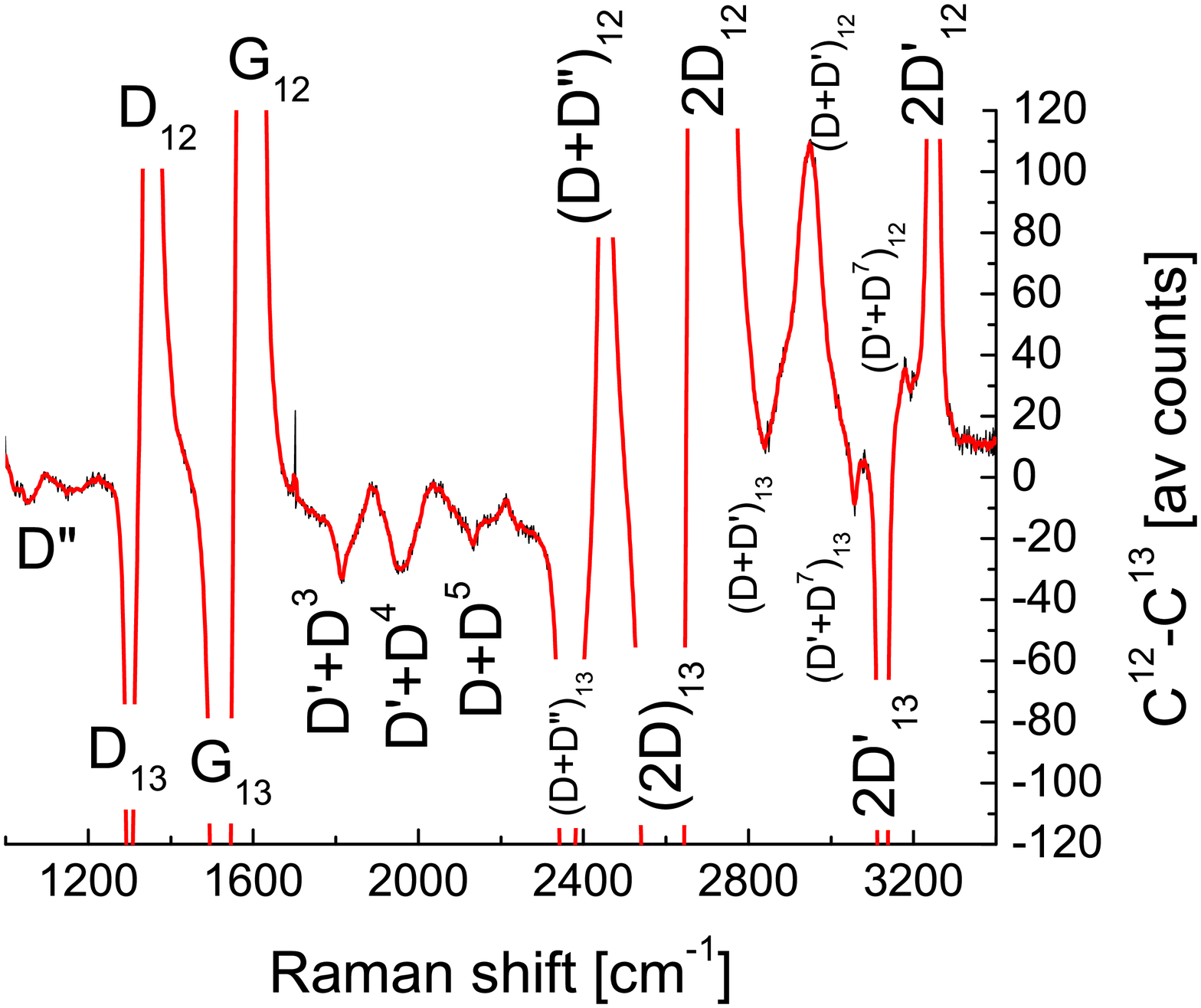}}
\caption{Difference between the average photon counts for C$^{12}$-graphene and C$^{13}$-graphene as an effective way to differentiate substrate and surface effects from the intrinsic graphene Raman spectrum. The red line is a smoothed average of the data (black line).} %
\vspace{-1cm}
\label{difference}%
\end{center}
\end{figure}

Since the substrate can affect the Raman spectrum, taking the difference between the spectra of the two isotopes allows to more clearly identify peaks which only correspond to graphene even in the presence of substrate induced peaks. For the rest of the analysis, which relies heavily on the determination of the peak positions, we used a combination of methods. For the large peaks, such as the 2D, 2D', G and D peaks, we fitted each peak after subtracting the substrate spectrum with a Lorentzian for every spot on the sample, before determining the mean and standard deviation of each peak. For very weak peaks this is difficult to do and we then only fitted the peaks after averaging over all spots. We then compared the peak positions obtained when subtracting only the average substrate spectrum with the peak positions obtained by subtracting the two average spectra of the two isotopes. Only if both match, we validate the peak positions. This is particularly important for the graphene peak positions below 1200 cm$^{-1}$, where the substrate induces large Raman peaks as shown in figures \ref{spectrum},\ref{difference}. It is worth noting that the peaks labeled D$^3$, D$^4$ and D$^5$ in figure \ref{difference} were not used in the determination of the phonon bandstructure discussed below, because of the weakness of the peaks. Instead, we used the double resonance peaks labeled D'+D$^3$, D'+D$^4$, and D+D$^5$, which are more prevalent.

\begin{figure}[ptb]
\begin{center}
\vspace{2cm}\includegraphics[width=3.5in]{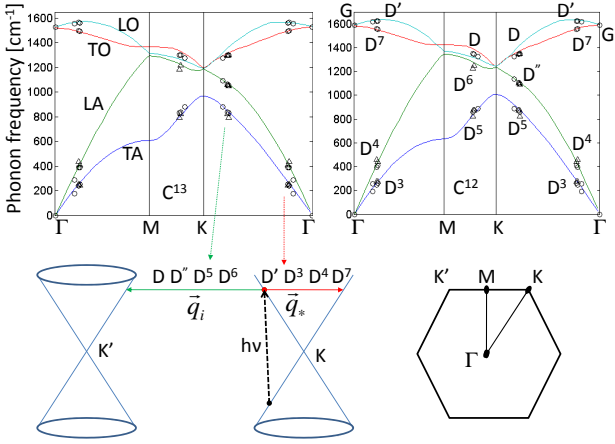}
\vspace{-3cm}
\caption{The top two graphs show the determined phonon energies at the corresponding wavenumbers superposed onto the theoretical BS taken from \cite{Mauri11} for the C$^{13}$ (left - scaled by $\sqrt{12/13}$) and C$^{12}$ (right) samples. In the top left we have labeled the 4 in-plane phonon branches, whereas the corresponding Raman lines are identified in the top right graph. The different phonon processes are illustrated in the bottom left, where $\vec{q}_*$ corresponds to an intra-valley phonon and $\vec{q}_i$ to an inter-valley phonon within the Brillouin zone depicted in the bottom right.}%
\vspace{-1cm}
\label{BS}%
\end{center}
\end{figure}

To understand the different Raman lines, we can divide them in three main processes: Process 1 (G-line), involves a phonon creation at the $\Gamma$ point, with zero momentum and hence the electron does not move in the Brillouin zone. This corresponds to the G line at $\sim$ 1586cm$^{-1}$ and it is the only Raman process involving a $\Gamma$ phonon, which leads to a single point on the band structure (BS) diagram at the $\Gamma$ point in figure \ref{BS}.

Processes 2 (intra-valley), are the ones involving a phonon emission when the excited electron scatters within its valley (K or K'). For processes between opposite sides of the cones, this leads to a phonon or electron wavenumber given by $q=\pi c/\lambda v_F$ assuming a linear electron dispersion $\epsilon=\hbar v_F |k|$ close to the K and K' points, where $c$ is the speed of light, $v_F\simeq 10^6$ms$^{-1}$ the Fermi velocity, $\lambda$ the incident Raman laser wavelength, and $k$ the electron wavenumber. These processes can involve TA phonons (D$^3$ Raman line), LA phonons (D$^4$ Raman line), TO phonons (D$^7$ Raman line) or LO phonons (D' Raman line) at momentum $q$ away from the $\Gamma$ point (labeled $\vec{q}_*$ in figure \ref{BS}.

Processes 3 (inter-valley) involve a phonon emission for an inter-valley (K to K') scattered electron. Assuming the electron scatters from opposite sides of the Dirac cones (outer processes, which are depicted in the K-M diagram), the corresponding wavenumber of the emitted phonon is K-K'+$q$. If the electron scatters between the inner sides (inner processes, which are depicted in the $\Gamma$-K diagram), the wavenumber is K-K'-$q$ (labeled $\vec{q}_i$ in figure \ref{BS}). Both interactions, K-K'$\pm q$ lead to Raman lines $q$ away from the K-point. These processes can involve TA phonons (D$^5$ Raman line), LA phonons (D" and D$^6$ Raman lines), and TO or LO phonons (D Raman line). The D Raman line is usually attributed to the TO branch \cite{Mafra07,Mauri11}.

\begin{figure}[ptb]
\begin{center}
\vspace{4cm}\includegraphics[width=3.5in]{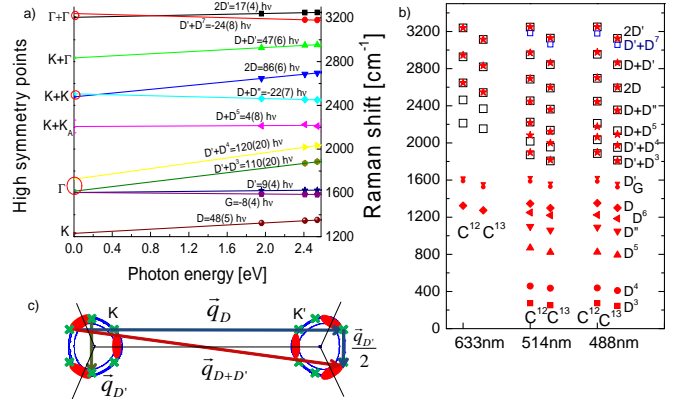}
\vspace{-4cm}
\caption{a) shows the laser excitation dependence of the Raman lines and their extrapolation to zero energy corresponding to the high symmetry points of the Brillouin zone. The values indicate the energy dependence of the Raman lines in units of cm$^{-1}/eV$ (the errors are in brackets). b) shows the equivalence between the energies of the double resonance processes (open squares) and the ones obtained by summing the corresponding single phonon processes (full red symbols). The missing data points, particularly for 1.9eV, reflects that some peaks are too weak for the peak position to be extracted. c) illustrates the mechanism attributed to the D+D' line. The green crosses on the electronic equi-energy lines (in blue and for laser excitations of 1.9, 2.4 and 2.5eV) correspond to the possible intra-valley processes, whereas red dots show the possible inter-valley processes.}%
\label{D+D'}%
\end{center}
\end{figure}

The various Raman lines can then be mapped onto the BS diagram as shown in figure \ref{BS}. However, some of the lines described above are very weak and barely visible, even when using the C$^{13}$-C$^{12}$ averaging method. This is the case for the lines D$^3$, D$^4$, D$^5$, D", and D$^7$, where we used their double resonance partners instead, in order to determine their precise energy. Double resonance processes, were described by Thomsen and Reich for graphite and involve two phonons emitted by the photon excited electron \cite{Reich00b}. The most dramatic example is the well studied 2D line, which is often used to determine the number of graphene layers in a given sample \cite{ferrari06}. The corresponding first order D-line is much weaker, since it requires the electrons to scatter back across to the other valley, which is only possible in the presences of short ranged impurities. Hence, while the D-line amplitude increases strongly in the presence of scattering impurities, the 2D peak remains the strongest Raman line for graphene monolayers. The 2D'-line is very similar (but weaker), except that it involves only intra-valley scattering. Both the 2D and 2D' processes involve two phonons with the same energy but opposite momentum. However, there are also a number of phonon pair processes which involve opposite momentum but with different energies. These are the combination process such as D'+D$^3$, D'+D$^4$, D+D$^5$, D+D", and D'+D$^7$. They are typically stronger than their single phonon counterparts because they do not involve an impurity scattering of the electron.

To show that we can indeed use these double resonance process in order to determine more precisely the Raman lines, we compared the frequencies corresponding to $\omega_{D^\nu+D^\mu}$ (labeled D$^\nu$+D$^\mu$ for a double phonon process) with $\omega_{D^\nu}+\omega_{D^\mu}$, where D$^\nu$ and $D^\mu$ are determined by the corresponding single phonon processes. We could do this for most of the observed double resonance Raman lines, which yields $\omega_{D^\nu+D^\mu}\simeq\omega_{D^\nu}+\omega_{D^\mu}$, within experimentally accuracy for most lines as shown in figure \ref{D+D'} and table \ref{table}. This justifies the use of the stronger second order processes to determine a more precise location of the Raman line and map it onto the corresponding single phonon BS point in figure \ref{BS}. A notable exception is the line labeled D+D', discussed below, which does not correspond to processes described above, since it involves the combination of an intra-valley phonon (D') and an inter-valley phonon (D).

\begin{table}
\begin{tabular}{|c|p{1.35cm}p{1.5cm}|p{1.25cm}p{1.35cm}p{1.75cm}|}
\hline
 & $\frac{\omega_{2D}}2-\omega_D$ & $\frac{\omega_{2D'}}2-\omega_{D'}$ & $\omega_{D}+\omega_{D'}$ $-\omega_{D+D'}$ & $\omega_D+\omega_G $ $-\omega_{D+D'}$ & $\omega_D+\frac{\omega_{D}+\omega_G}2$ $-\omega_{D+D'}$\\
\hline
C$^{13}_{488}$ & -2.5$\pm$ 0.3 & -0.3 $\pm$ 0.4 & 18$\pm$ 2 & -21$\pm$ 2 & -1.5 $\pm$ 1.8\\
C$^{12}_{488}$ & -2.3$\pm$ 0.1 & -0.1$\pm$0.2 & 19$\pm$ 1 & -20$\pm$ 1 & -0.6 $\pm$ 1\\
C$^{13}_{514}$ & -1.8$\pm$ 0.2 & -1.4$\pm$ 0.3 & 18$\pm$ 2 & -18$\pm$ 2 &  0.1 $\pm$ 1.4\\
C$^{12}_{514}$ & -1.7$\pm$ 0.3 & 0.5$\pm$ 0.3 & 18$\pm$ 1 & -19$\pm$ 1 & -0.4 $\pm $ 1.1\\
C$^{13}_{633}$ & -1.3$\pm$ 0.9 & -1$\pm$ 0.7 & 9$\pm$ 10 & -23$\pm$ 10 & -6.7 $\pm $ 10\\
C$^{12}_{633}$ & -1.1$\pm$ 0.2 & -0.9$\pm$ 0.8 & 16$\pm$ 6 & -14$\pm$ 6 & 0.9 $\pm $ 6\\
\hline
\end{tabular}
\caption{List of the difference of some Raman lines for different laser wavelengths and for both isotopes. The errors correspond to the standard error of the mean except for the lines involving the D+D' process at a laser wavelength of 633nm, where we could only fit the average over all spots. All numbers are in units of cm$^{-1}$.}
\label{table}
\end{table}

It turns out that $\omega_{D}+\omega_{D'}>\omega_{D+D'}>\omega_{D}+\omega_{G}$ as shown in table \ref{table}. We can shine light on the D+D' process by considering the situation illustrated in figure \ref{D+D'}. The D'-line (intra-valley) was shown to predominantly occur at opposite sides of the Dirac cone along the three K-M directions (indicated by small crosses in figure \ref{D+D'} and labeled $\vec{q}_{D'}$). The D-line (inter-valley) was shown to occur mainly between opposite ends of the K and K' cones along elongated constant energy surfaces \cite{Mauri11,Narula08} (indicated by red elongated dots in figure \ref{D+D'}). The wavenumber $\vec{q}_{D}$ connects two of these ends illustrating the outer process. This electron can then either scatter back with the help of an impurity to its original momentum (a standard D-line process) or it can first emit a $\eta\vec{q}_{D'}$ phonon to  scatter the electron across the valley before it recombines through an impurity scattering like the D-line process. The possible values for $\eta$ are 0, $\frac{1}2$ or 1 for scattering along the K-M direction (depicted in figure \ref{D+D'} is the case $\eta=\frac{1}2$). On average we have $\eta\simeq \frac{1}2$, hence the electron loses an energy $\omega_{\frac{1}2 q_{D'}}+\omega_{q_D}$, which we labeled D+D' and therefore yields the energy $\frac{\omega_{D'}+\omega_{G}}2+\omega_D$, tabulated in table \ref{table} and corresponding precisely to $\omega_{D+D'}$. A similar interaction can happen for the inner process, which would lead to the same energy loss. The D+D' line is therefore consistent with the double phonon process involving a both an inter-valley phonon and a intra-valley phonon.

In figure \ref{D+D'} we show the dependence of the various double resonance lines on the energy of the laser. This further constrains the assignment of the Raman peaks to the different phonon modes. Indeed, The 2D' and D'+D$^7$ lines are expected to merge to twice the G-line at the $\Gamma$ point at zero laser energy, whereas the D'+D$^3$, D'+D$^4$, for example should merge to the G-line at zero energy at the $\Gamma$ point. This is indeed what happens, further confirming the correct assignments of all the observed peaks.

Thanks to the high precision in determining peak positions of the strong D and 2D peaks, we can see a very small difference (but statistically significant) between $\frac{\omega_{2D}}2$ and $\omega_D$ ($\sim 2$ cm$^{-1}$ as shown in table \ref{table}. This is possibly due to the relative importance of the inner versus outer processes, which differ by about 5 cm$^{-1}$ according to Venezuela et al. \cite{Mauri11}, since the momentum dependence of the scattering rates of the electron-phonon process (important for the 2D line) are different from rates of the electron-impurity process only relevant to the D line. In contrast, the difference between $\frac{\omega_{2D'}}2$ and $\omega_D'$ is negligible, which is consistent with the theoretically obtained orientation dependence of only 1 cm$^{-1}$ \cite{Mauri11}. This also explains the narrower 2D' peak in comparison to the 2D peak (see fig. \ref{spectrum}).

Summarizing, we identified several new Raman peaks in graphene, using a combination of isotopes and Raman laser energies on large scale CVD grown graphene, which allowed us to do extensive configurational averaging, without increasing the laser power. In particular, we related the peak labeled D+D' to a double phonon process involving both an inter- and intra-valley phonon. The large number of identified peaks allowed us to map out a general in-plane phonon band structure of graphene using the excitation energy dependence of the various Raman lines.

$^*$ Also at the Institute for Microstructural Sciences M23A,
National Research Council of Canada, Ottawa, Ontario K1A 0R6, Canada.

\end{document}